\documentclass[aps,prl,twocolumn,superscriptaddress]{revtex4}
\usepackage{graphicx}
\usepackage{hyperref}

\def\beq{\begin{equation}}
\def\eeq{\end{equation}}
\def\bea{\begin{eqnarray}}
\def\eea{\end{eqnarray}}

\def\bwt{\begin{widetext}}
\def\ewt{\end{widetext}}

\nocite{*}

\usepackage{hyperref}

\begin{document}

\title{Does Planck mass run on the cosmological horizon scale?}
\author{Georg Robbers}\email{g.robbers@thphys.uni-heidelberg.de}
\affiliation{Institut f\"ur Theoretische Physik, Philosophenweg 16,
69120 Heidelberg, Germany}
\author{Niayesh Afshordi}\email{nafshordi@cfa.harvard.edu}\affiliation{Institute for
  Theory and Computation,
Harvard-Smithsonian Center for Astrophysics, MS-51, 60 Garden
Street, Cambridge, MA 02138, USA}\affiliation{Perimeter Institute
for Theoretical Physics, 31 Caroline St. N., Waterloo, ON, N2L 2Y5,
Canada}
 \author{Michael Doran}\email{M.Doran@thphys.uni-heidelberg.de}
\affiliation{Institut f\"ur Theoretische Physik, Philosophenweg 16,
69120 Heidelberg, Germany}
\date{\today}
\preprint{astro-ph/yymmnnn}
\begin{abstract}

Einstein's theory of general relativity, which contains a universal
value of the Planck mass, has been so far successfully invoked to
explain gravitational dynamics from sub-millimeter scales to the
scale of the cosmological horizon. However, one may envisage that in
alternative theories of gravity, the effective value of the Planck
mass (or Newton's constant), which quantifies the coupling of matter
to metric perturbations, can run on the cosmological horizon scale.
In this letter, we study the consequences of a glitch in the Planck
mass from sub-horizon to super-horizon scales. We first give three
examples of models that naturally exhibit this feature, and then
show that current cosmological observations severely constrain this
glitch to less than 1.2\%. This is the strongest constraint to date,
on {\it natural} (i.e. non-fine-tuned) deviations from Einstein
gravity on the cosmological horizon scale.

\end{abstract}
\maketitle

\section{Introduction}
The Einstein theory of gravity (or General Relativity) is among the
most successful theories in physics. Despite its simple mathematical
structure, and having only a single constant, it has been successful
in explaining the cosmological observations on the horizon scale
($\sim 10^{25-28} {\rm cm}$), down to the planetary/lunar dynamics
on the solar system scales ($\sim 10^{9-15} {\rm cm}$), and even
laboratory tests of the inverse square law on the sub-millimeter
scales (see \cite{Adelberger:2003zx} for an overview).

These tests, as well as a slew of other astrophysical observations,
severely constrain any alternative to the Einstein theory of
gravity.
Nevertheless, deviations from Einstein gravity are
 expected, just based on theoretical grounds. Einstein gravity is a
classical theory, which does not have a well-defined quantization,
and thus a more complete theory of gravity (such as string theory)
is necessary to describe gravitational interaction at high energies
(or small scales).
However, a full theory of quantum gravity is not
necessary in most models (with notable exceptions), unless we want
to study interactions at very small scales (Planck length $\sim
10^{-33} {\rm cm}$), which are far from the range accessible in
terrestrial experiments or astrophysical processes.

Deviations from Einstein gravity have also been suggested on purely
phenomenological grounds, in particular to explain the rotation
curves of galaxies (as a replacement for dark
matter)\cite{Milgrom:1983ca,Bekenstein:2004ne}, or the
discovery of the apparent acceleration of cosmic expansion (as a
replacement for dark energy/cosmological constant)
\cite{Carroll:2003wy,Deffayet:2000uy}. However, the evidence for any
such deviation (rather than simply exotic matter/energy
components), is far from conclusive.

A customary way to quantify deviations from Einstein gravity on
small scales and in the weak field limit, is through introducing a
Yukawa fifth force modification to the inverse square law, where the
gravitational potential energy takes the form: \beq V(r)=
-G\frac{m_1m_2}{r}\left(1+\alpha e^{-r/\lambda}\right),\label{isl}
\eeq where $G$ is the Newton's constant, $m_1$ and $m_2$ are the
masses of (point-like) gravitating objects, while $\alpha$ and
$\lambda$ quantify the strength and the scale of the new
interaction, respectively. In this model, the effective Newton's
constant smoothly goes from $G$ on large scales ($r\gg \lambda$) to
$G(1+\alpha)$ on small scales ($r \ll \lambda$). Current
experimental and observational constraints severely limit $\alpha$
in the range $ 10^{-1} {\rm cm} < \lambda < 10^{16} {\rm cm}$ (see
\cite{Adelberger:2003zx} for an overview).

In this letter, we investigate the possibility of a similar glitch
in the Newton's constant (or Planck mass) on the scale of the
cosmological horizon, or the Hubble radius ($\lambda \sim c/H \sim
10^{28} {\rm cm}$). In our case, the scale $\lambda$ will not be a
physical constant of the theory, but rather an emergent scale in the
theory, as a consequence of an effective change in the background
geometry, from flat Minkowski space on small scales, to the expanding
Friedmann-Robertson-Walker background on large scales.

We start by defining an effective Planck mass, and then give a few
examples of the theories which contain a glitch in their effective
Planck masses on the horizon scale. We will then investigate the
cosmological consequences of such a glitch for structure formation,
and the cosmic microwave background, and provide a limit based on
current cosmological observations.

\section{Effective Planck Mass}

The Planck mass, $M_p$, quantifies the strength of coupling between
the space-time metric, and the energy-momentum of matter in the
Universe. In terms of the Einstein equation: \beq G^\mu_{\nu} =
M^{-2}_p T^\mu_{\nu}, \label{einstein}\eeq where $G^\mu_\nu$ and
$T^\mu_\nu$ are the Einstein and the total energy-momentum tensors
respectively. Notice that in our notation, $M_p = (8\pi G)^{-1/2}
\simeq 2.44 \times 10^{18} {\rm GeV}$, where $G$ is Newton's
gravitational constant, and we have used natural units ($\hbar = c =
1$).

Even though the Planck mass is a constant of the Einstein theory of
gravity, possible deviations from Einstein gravity, or
alternatively, other energy components that are not accounted for in
the total energy momentum tensor, $T_{\mu\nu}$, may lead to an
effective (or dressed) Planck mass that could run with time and/or
the energy/length scale of the interactions. A possible definition for
an effective Planck mass may come by perturbing the Einstein
constraint (or $G^0_0$) equation: \beq M^{-2}_{p, {\rm eff}} \equiv
\frac{\delta G^0_0}{\delta T^0_0}, \label{mpt}\eeq which reduces to
the Poisson equation around a Minkowski background (or on
sub-horizon scales). However, Eq. (\ref{mpt}) can mix different
scales, as it involves the ratio of two variable functions.
Moreover, this definition may become ill-defined if $\delta T^0_0$
crosses zero. Instead, we are going to adopt a more practical
definition: \beq M^{-2}_{p, {\rm eff}}(|{\bf k}|) \equiv
\frac{\langle\delta G^0_{0,{\bf k}} \delta T^{0*}_{0,{\bf k}}\rangle
}{\langle \delta T^0_{0,{\bf k}} \delta T^{0*}_{0,{\bf k}} \rangle
}, \eeq where $\delta T^0_{0,{\bf k}}$ and $\delta G^0_{0,{\bf k}}$
are the spatial Fourier transforms of $\delta T^0_0$ and $\delta
G^0_0$ on a given spatial hypersurface.

While this definition has the benefit of separating different
physical scales, we have introduced an explicit gauge-dependence
through the choice of a particular spatial hypersurface. On small
(sub-horizon) scales \mbox{($k \gg H$)}, this gauge dependence is not
important, as \mbox{Eq. (\ref{mpt})} reduces to the Poisson equation,
and we recover Newtonian gravity. In other words, the difference
between $M_{p,{\rm eff}}$ in different (physical) gauges is $\sim
(k/H)^{-2}$, on sub-horizon scales.

On super-horizon scales ($k \ll H$), gauge transformations can
change the effective Planck mass, only if the Planck mass associated
with the background expansion is different from the Planck mass
associated with the perturbations, i.e. as long as: \beq
M^{-2}_{p,{\rm IR}} = \frac{\dot{G}^0_0}{\dot{T}^0_0} = \frac{\delta
G^0_0}{\delta T^0_0}, \eeq the effective Planck mass is
gauge-invariant on super-horizon scales. This condition can
naturally result from the assumption of adiabatic initial
conditions, which asserts that, up to a time shift, causally
disconnected patches of the Universe experience identical histories.
Therefore, we see that, at least for {\it adiabatic} initial
conditions, the gauge dependence of our definition of the effective
Planck mass may only become important as modes cross the horizon.

From here on, we will refer to the cosmological sub-horizon ($k \gg
H$) and super-horizon ($k \ll H$) scales as the UV and IR scales,
respectively, which have their respective values of the effective
Planck mass, $M_{p,{\rm UV}}$ and $M_{p,{\rm IR}}$. In the language
of Eq. (\ref{isl}), the UV-IR mismatch can be parametrized by the
dimensionless $\alpha$ parameter: \beq M^2_{p,{\rm IR}} =
M^2_{p,{\rm UV}} (1+\alpha). \label{alpha_def}\eeq

\section{Three Examples}

As an example, let us consider the quadratic Cuscuton action
\cite{Afshordi:2006ad}: \beq \label{cus_action}S_Q = \int d^4x
\sqrt{-g} \left(\mu^2
\sqrt{|\partial^\mu\varphi\partial_\mu\varphi|}-\frac{1}{2}m^2\varphi^2\right),
\eeq where $\varphi$ is a scalar field, and $\mu$ and $m$ are
constants of theory with the dimensions of energy.

If we consider Cuscuton as a part of the gravitational action (and
so do not include it in the energy-momentum tensor), the effective
Planck mass takes the form: \beq M^{-2}_{p, {\rm eff}}(k) =
\frac{\delta\rho_{Q} +\delta\rho_{m}}{\delta\rho_{m}}.
\label{cus_mpeff} \eeq Using the solution to the field equation in
the Longitudinal gauge, obtained in \cite{Afshordi:2007yx}, we find
that: \beq M^2_{p,{\rm eff}} \simeq M^2_{p,{\rm UV}} -{ 3\mu^4\over
2m^2}
\left(1+\frac{k^2}{3H^2}\right)^{-1}\left(1-\frac{k^2}{3\dot{H}}\right)^{-1},
\eeq to the lowest order in $\mu$, in a flat matter-dominated
Universe.

As we noted above, the exact k-dependence of $M_{p,{\rm eff}}$ will
depend on the choice of gauge, but the IR limit of the effective
Planck mass: \beq M^2_{p,{\rm IR}} = M^2_{p,{\rm UV}} -{ 3\mu^4\over
2m^2}, \label{cus_ir}\eeq is set by the Friedmann equation
\cite{Afshordi:2006ad}, and is thus gauge-invariant \footnote{In
fact, Eq. (\ref{cus_ir}) holds exactly for quadratic Cuscuton,
independent of the equation of state of the rest of the energy
components of the Universe.}.

A very similar behavior can be seen in the dynamics of a canonical
scalar (or quintessence) field with a simple exponential potential
\cite{Ferreira:1997hj}: \beq V(\varphi) = M^4_p e^{-\kappa
\varphi/M_p}. \eeq It is easy to see that, for a fixed background
equation of state, the energy density of the field, asymptotically,
reaches a constant fraction of the energy density of the Universe.
In particular, for a flat matter-dominated cosmology, this fraction
is: \beq \label{eq::omega_quint} \Omega_{\varphi} =
\frac{3}{\kappa^2}, \eeq which
translates to a glitch in the effective Planck mass on the horizon
scale: \beq \label{eq::Mp_quint} M^2_{p,{\rm IR}} =
M^2_{p,UV}\left(1-\frac{3}{\kappa^2}\right), \eeq as quintessence
does not cluster on sub-horizon scales, and so $M_{p,{\rm eff}}$
reaches its fundamental value on small scales.

A third model that leads to a similar mismatch between the IR and UV
effective Planck masses has been introduced by Carroll and Lim
\cite{Carroll:2004ai}, and consists of a Lorentz-violating,
fixed-norm, time-like vector field $u^{\mu}$, with the Lagrangian:
\bea {\cal L}_u &=& -\beta_1
\nabla^{\mu}u^{\sigma}\nabla_{\mu}u_{\sigma}-\beta_2
\left(\nabla_{\mu}u^{\mu}\right)^2-\beta_3
\nabla^{\sigma}u^{\mu}\nabla_{\mu}u_{\sigma}\nonumber\\ &+&\lambda_u
(u^{\mu}u_{\mu}+m_u^2), \eea where $m_u$ is the norm of the vector
field, $\lambda_u$ is a Lagrange multiplier, and $\beta_i$'s are
dimensionless constants of the theory. In this model, the coupling
of $u^{\mu}$ to the metric renormalizes both the IR and UV values of
the effective Planck mass, so that: \beq M^2_{p,{\rm IR}} =
M^2_{p,{\rm UV}} + (2\beta_1+3\beta_2+\beta_3)m^2_u. \eeq It is
interesting to notice that, unlike the two scalar field models
discussed before,
the gravitational force is suppressed on large (super-horizon)
scales by the Lorentz-violating vector field \cite{Carroll:2004ai}.

We will next look at the observational consequences of a possible
glitch between the UV and IR effective Planck masses.

\section{Cosmological Constraints on the running of the Planck mass}
\begin{figure}
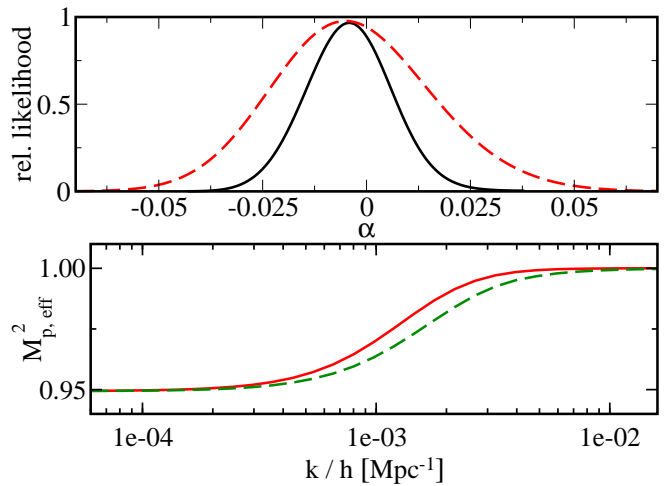

\includegraphics[width=\columnwidth]{constraints-alpha.eps}
\includegraphics[width=\columnwidth]{Mp2_eff.eps}
\caption{\textit{Top Panel:} Observational constraints on the UV/IR
Planck mass mismatch parameter, $\alpha$, from 3 years of WMAP data
alone \cite{Hinshaw:2006ia}(red, dashed line), and our compilation
(see the text) of cosmological observations (black straight line).
\textit{Bottom Panel:} Transition between the IR and UV regimes for
the effective Planck mass (in units of $M_{p,{\rm IR}}$) defined in
Eq.(\ref{cus_mpeff}) for the quadratic Cuscuton with $\alpha =
-0.05$ (red, straight line). The transition for a canonical scalar
field model ($c_s^2=1$) is depicted in green (dashed line). For
$c_s^2 \gtrsim 10$, the transitions virtually coincide with the
quadratic Cuscuton (for which $c_s^2 =\infty$). } \label{figure}
\end{figure}
The Hubble expansion rate in a flat homogenous cosmology is set by
the Friedmann equation: \beq H^2 = \frac{\rho_{\rm
tot}}{3M^2_{p,{\rm IR}}}. \label{Friedmann}\eeq As the present-day
cosmic density, $\rho_{\rm tot}$, is dominated by dark matter and
dark energy, which are only seen through their gravitational
effects, there is no way to find $M_{p,{\rm IR}}$ through measuring
the present-day Hubble constant. However, the energy density in the
radiation era is dominated by photons and neutrinos (for $T \lesssim
1~ {\rm MeV}$), which are better understood, as their energy density
(for three species of relativistic neutrinos) is fixed by the Cosmic
Microwave Background (CMB) temperature ($T = 2.728 \pm 0.004~ {\rm
K}$, \cite{Fixsen:1996nj}). Constraints on the expansion rate during
the radiation era (at $T\sim 0.1~{\rm MeV}$) then come from
comparing the Big Bang Nucleosynthesis predictions with the
cosmological observations of the light element abundances, which
correspondingly constrain the running of the Planck mass: $\alpha =
0.0 \pm 0.2$ ($95\%$ confidence level) \cite{Cyburt:2004yc}.

More interesting constraints can come from the study of cosmological
perturbations on small scales. Combining the continuity and Poisson
equations with Newton's 2nd law yields: \beq
\ddot{\delta}+2H\dot{\delta} = \frac{\bar{\rho}_{m}}{2M^2_{p,{\rm
UV}}}\delta, \eeq for the linear matter overdensity perturbations
$\delta$ \mbox{($=\delta\rho_m/\bar{\rho}_m$)} on small scales.
Combining this with the Friedmann equation (Eq. \ref{Friedmann}) in
the matter-dominated era, and the definition of $\alpha$ (Eq.
\ref{alpha_def}), we find: \beq \ddot{\delta}+2H\dot{\delta}-
\frac{3}{2}(1+\alpha)H^2\delta=0, \eeq which can be easily solved
(using $H=2/3t$, in the matter-dominated era). For $\alpha \ll 1$,
the growing mode behaves as: \beq \delta \propto
t^{\frac{2}{3}+\frac{2}{5}\alpha} \Rightarrow \Phi \propto
t^{\frac{2}{5}\alpha},\eeq where $\Phi$ is the Newtonian (or
longitudinal metric) potential.

Perturbation modes that are inside the horizon at the time of
matter-radiation equality will then all experience the same amount
of suppression or enhancement during the matter era. Since the scale
factor grows as $t^{2/3}$, this suppression/enhancement is roughly
by a factor of $z_{\rm eq}^{3\alpha/5} \simeq 1+5\alpha$, where
$z_{\rm eq} \simeq 3400$ is the redshift of matter-radiation
equality.

For the angular spectrum of CMB anisotropies, this will result in a
small change in the power on small scales, and also in a change of
the contribution from the Integrated Sachs-Wolfe (ISW) effect, as a
result of the decaying/growing Newtonian potential
\footnote{Presence of anisotropic stress in some modifications to
Einstein gravity, such as the Lorentz-violating vector field, can
change the predictions for the ISW effect. However, given the
current constraints on the anisotropic stress
\cite{Caldwell:2007cw}, this is unlikely to affect our results
significantly.}. For $\alpha>0$, there will be less power on the
large scale CMB power spectrum, whereas a negative $\alpha$ will
lead to an increase of power. Correspondingly, the acoustic peak of
the CMB power spectrum will be slightly shifted, due to the change
in the cosmic expansion history (see Fig.2 in
\cite{Afshordi:2007yx}, in which $\Omega_Q = -\alpha$). Therefore, a
possible running of the Planck mass on the horizon scale can be
constrained by CMB observations. We compute these constraints using
a modified version of cmbeasy \cite{Doran:2003sy} for the quadratic
Cuscuton model (see \cite{Afshordi:2007yx} for details). We find
that the 3-year CMB power spectrum of WMAP \cite{Hinshaw:2006ia}
constrains $\alpha$ to $-0.005 \pm 0.040$ (at 95\% confidence).

The impact of a UV/IR glitch in the effective Planck mass on
structure formation could be equally significant. The most prominent
effect is the change in the amplitude of the matter power spectrum
(in comparison to the CMB power) on small scales at late times. In
addition, modes entering the horizon at different times will be
suppressed or enhanced (depending on the sign of $\alpha$) by a
factor which depends on the time when they enter the horizon, as we
have seen above. As a result, the cold dark matter power spectrum
will also be tilted between the equality and present-day horizon
scales.

Hence, the amplitude of a possible UV/IR mismatch of the effective
Planck mass can be also constrained by observations of large scale
structure. We use the latest data from the distribution of luminous
red galaxies from the Sloan Digital Sky Survey (SDSS)
\cite{Tegmark:2006az} (marginalizing over bias). We also include
constraints on the cold dark matter power spectrum from observations
of the Lyman-$\alpha$ forest \cite{Seljak:2006bg}. Even though this
data extends into the mildly non-linear regime of the power
spectrum, we expect non-linear effects (of a non-vanishing $\alpha$)
to be of little importance here, in particular because the bounds on
$\alpha$ are already rather tight from CMB alone. Adding the results
from Supernovae Ia observations \cite{Riess:2006fw} as well as the
observation of the baryon acoustic oscillations (BAO)
\cite{Eisenstein:2005su} to this large scale structure data (in
order to decrease degeneracy with other cosmological parameters),
and the data from 3 years of WMAP \cite{Hinshaw:2006ia}, we find the
UV/IR mismatch parameter, $\alpha$, to be tightly constrained to
$-0.004 \pm 0.021$ (95\% CL) by our complete set of current
observational data (see Fig. \ref{figure}).

\section{Discussions}
One may wonder if our constraints on $\alpha$ may depend on the
specific model that yields the running of the effective Planck mass.
 Fig. \ref{figure} compares the UV-IR transition of
$M^2_{p,{\rm eff}}$ (in longitudinal gauge) for the quadratic
Cuscuton and the exponential scalar field models in the matter era.
We see that for the canonical scalar field model the transition is
shifted to  slightly smaller scales by $\sim 30\%$.
However, the effect on the
bounds on a UV/IR glitch that were computed above is marginal. For
the scalar field model with $c_s^2=1$, the bounds on $\alpha$ are
only slightly relaxed, namely
to $-0.004\pm 0.024$ at the $95\%$ confidence level. Therefore, we
conclude that the constraints on $\alpha$ are insensitive to the
details of UV/IR transition, since most of the observable
consequences of the mismatch occur on small sub-horizon scales.

What about an arbitrary redshift evolution of the running factor,
$\alpha(z)$? While this is, in principle possible, and in line with
the current efforts to quantify the redshift evolution of dark
energy, we should point out that any such evolution would require
introducing an {\it ad hoc} macroscopic scale (coincident with the
present-day horizon) into the theory. Indeed, this is the same
fine-tuning problem that most dark energy or modified gravity
theories (with non-trivial late-time dynamics) suffer from. In lieu
of any such scale, the cosmological horizon is the only macroscopic
scale in the problem that could control the running of gravitational
coupling constants. Therefore, a constant $\alpha$ is the only
natural result of a non-trivial {\it microscopic} physics in the
gravity theory.

To summarize, in this letter, we have studied the running of the
Planck mass (or Newton's constant) on the cosmological horizon
scale, as a possible modification of Einstein gravity. We have first
discussed three physical models that naturally exhibit this running.
We then considered observable consequences of this running and found
out that any mismatch between UV and IR Planck masses (Newton's
constants) is severely constrained to less than 1.2\% (2.4\%) at the
95\% confidence level. While future cosmological observations are
likely to strengthen this bound by an order of magnitude over the
next decade, the expected magnitude of such a running/glitch in
well-motivated extensions to Einstein gravity is yet to be
determined.

It is our pleasure to thank Daniel Chung, Ghazal Geshnizjani, Justin
Khoury, Christof Wetterich, and Matias Zaldarriaga for helpful
discussions and useful comments. NA also wishes to thank the
hospitality of the Physics department at the University of
Wisconsin-Madison, during the course of this work.

\bibliographystyle{utphys_na}
\bibliography{planck}

\end{document}